\documentclass[submission, Phys]{SciPost}

\hypersetup{
    colorlinks,
    linkcolor={red!50!black},
    citecolor={blue!50!black},
    urlcolor={blue!80!black}
}

\usepackage[bitstream-charter]{mathdesign}
\usepackage{bm}

\DeclareSymbolFont{usualmathcal}{OMS}{cmsy}{m}{n}
\DeclareSymbolFontAlphabet{\mathcal}{usualmathcal}

\begin{document}

\begin{center}{\Large \textbf{
Laser excitation of the  1s-hyperfine {transition}  in muonic hydrogen\\
}}\end{center}

\begin{center}
P. Amaro\textsuperscript{1$\star$},
A. Adamczak\textsuperscript{2},
M. Abdou Ahmed\textsuperscript{3},
L. Affolter\textsuperscript{4},
F. D. Amaro\textsuperscript{5},
P. Carvalho\textsuperscript{1},
T.-L. Chen\textsuperscript{6},
L. M. P. Fernandes\textsuperscript{5},
M. Ferro\textsuperscript{1},
D. Goeldi\textsuperscript{4},
T. Graf\textsuperscript{3},
M. Guerra\textsuperscript{1},
T. W. H\"{a}nsch\textsuperscript{7,8}, \\
C. A. O.  Henriques\textsuperscript{5},
Y.-C. Huang\textsuperscript{6},
P.~Indelicato\textsuperscript{9},
O.~Kara\textsuperscript{4},
K.~Kirch\textsuperscript{4,10},
A.~Knecht\textsuperscript{10},\\
F. Kottmann\textsuperscript{4,10},
Y.-W. Liu\textsuperscript{6},
J. Machado\textsuperscript{1},
M. Marszalek\textsuperscript{10},
R. D. P. Mano\textsuperscript{5},
C. M. B. Monteiro\textsuperscript{5},
F. Nez\textsuperscript{9},
J. Nuber\textsuperscript{4,10},
A. Ouf\textsuperscript{11},
N. Paul\textsuperscript{9},
R. Pohl\textsuperscript{11, 12},
E. Rapisarda\textsuperscript{10},
J. M. F. dos Santos\textsuperscript{5}, \\
J. P. Santos\textsuperscript{1},
P.~A.~O.~C. Silva\textsuperscript{5},
L. Sinkunaite\textsuperscript{10},
J.-T. Shy\textsuperscript{6},
K. Schuhmann\textsuperscript{4},
S. Rajamohanan\textsuperscript{11},
A. Soter\textsuperscript{4},
L. Sustelo\textsuperscript{1},
D. Taqqu\textsuperscript{4,10},
L.-B. Wang\textsuperscript{6},
F. Wauters\textsuperscript{12},
P. Yzombard\textsuperscript{9},
M. Zeyen\textsuperscript{4},
A. Antognini\textsuperscript{4,10$\dagger$},
\end{center}

\begin{center}
{\bf 1} Laboratory of Instrumentation, Biomedical Engineering and Radiation Physics (LIBPhys-UNL), Department of Physics, NOVA School of Science and Technology, NOVA University Lisbon, 2829-516 Caparica, Portugal
\\
{\bf 2} Institute of Nuclear Physics, Polish Academy of Sciences, PL-31342 Krakow, Poland
\\
{\bf 3} Institut f\"ur Strahlwerkzeuge, Universit\"at Stuttgart, 70569 Stuttgart, Germany
\\
{\bf 4} Institute for Particle Physics and Astrophysics, ETH Zurich, 8093 Zurich, Switzerland
\\
{\bf 5} LIBPhys-UC, Department of Physics, University of Coimbra, \mbox{P-3004-516 Coimbra, Portugal}
\\
{\bf 6} Department of Physics, National Tsing Hua University, Hsinchu 30013, Taiwan
\\
{\bf 7} Ludwig-Maximilians-Universit\"at, Fakult\"at f\"ur Physik, 80799 Munich, Germany
\\
{\bf 8} Max Planck Institute of Quantum Optics, 85748 Garching, Germany
\\
{\bf 9} Laboratoire Kastler Brossel, Sorbonne Universit\'{e}, CNRS, ENS-Universit\'e PSL, Coll\`{e}ge de France, 75005 Paris, France
\\
{\bf 10} Paul Scherrer Institute, 5232 Villigen–PSI, Switzerland
\\
{\bf 11} Institut f\"ur Physik, QUANTUM, Johannes Gutenberg-Universit\"at Mainz, 55099 Mainz, Germany
\\
{\bf 12} PRISMA+ Cluster of Excellence and Institute of Nuclear Physics, Johannes Gutenberg-Universit\"at Mainz, 55099 Mainz, Germany \\

~

${}^\star${\small \sf pdamaro@fct.unl.pt}
${}^\dagger${\small \sf aldo.antognini@psi.ch}
\end{center}

\begin{center}
\today
\end{center}


\section*{Abstract}
{\bf
The CREMA collaboration is pursuing a measurement of the  ground-state hyperfine splitting (HFS) in muonic hydrogen ($\mu$p) with 1 ppm accuracy by means of pulsed laser spectroscopy to determine the two-photon-exchange contribution with $2\times10^{-4}$ relative accuracy.   In the proposed experiment, the  $\mu$p atom  undergoes a laser excitation from the singlet hyperfine state to the triplet hyperfine state, {then} is quenched back to the singlet state  by an inelastic collision with a H$_2$ molecule. The resulting increase of kinetic energy after the collisional deexcitation is used as a signature of a successful laser transition between hyperfine states.  In this paper, we calculate the combined probability that a $\mu$p atom initially in the singlet hyperfine state undergoes a laser excitation to the triplet state followed by a collisional-induced deexcitation back to the singlet state.  This combined probability has been computed using the optical Bloch equations including the inelastic and elastic collisions. Omitting the decoherence effects caused by {the laser bandwidth and }collisions would overestimate the transition probability by more than  a factor of {two in the experimental conditions. Moreover,}  we also account for Doppler effects and provide the matrix element, the saturation fluence, the elastic and inelastic collision rates for the singlet and triplet states, and the resonance linewidth. This calculation thus quantifies one of the key unknowns of the HFS experiment, leading to a precise definition of the requirements for the laser system and to an optimization of the hydrogen gas target where $\mu$p is formed and the laser spectroscopy will occur. 
}

\vspace{10pt}
\noindent\rule{\textwidth}{1pt}
\tableofcontents\thispagestyle{fancy}
\noindent\rule{\textwidth}{1pt}
\vspace{10pt}

\section{Introduction}
\label{intro}

Highly accurate measurements of atomic transitions can be used
as precise probes of low-energy properties of nuclei.
As the Bohr radius of hydrogen-like atoms decreases with increasing
orbiting particle mass, muonic atoms (atoms formed
by a negative muon and a nucleus) have an enhanced sensitivity to nuclear
structure effects~\cite{Eides2001, Karshenboim2005, Pachucki1999,
  Pohl2010, Antognini2013b, Antognini2013c, Peset2021}.
The Charge Radius Experiments with Muonic Atoms (CREMA) collaboration in recent years
has performed laser spectroscopy of the $2s-2p$ (Lamb shift) transitions in muonic hydrogen ($\mu$p) \cite{Pohl2010, Antognini2013c},
muonic deuterium ($\mu$d) \cite{Pohl2016} and muonic helium {($\mu^4$He$^+$)} \cite{Krauth2021} and extracted the
corresponding nuclear charge radii with an unprecedented accuracy.
The impact of the $\mu$p measurements on beyond-standard-model searches, on precision atomic physics, and  on the proton structure can be found in recent reviews~\cite{Gao2021, Karr2020, Carlson2015, Peset2021}. 
Along this line of research, the CREMA collaboration is
presently aiming at the measurement of the ground-state hyperfine
splitting (HFS) in $\mu$p with 1~ppm relative accuracy by
means of pulsed laser spectroscopy.

From the measurement of the HFS, precise information about the magnetic structure of the
proton can be extracted~\cite{Peset2017, Tomalak2017, Tomalak2019, Hagelstein2016, Karshenboim2014, Alarcon2020, Volotka2005, Dupays2003, Peset2014, Hagelstein2018, Carlson2011}. 
Specifically, by comparing the measured HFS transition frequency
with the corresponding theoretical prediction based on bound-state QED
calculations~\cite{Peset2017, Volotka2005, Dupays2003,
  Antognini2013b}, the two-photon-exchange contribution can be
extracted with {approximately} $2\times 10^{-4}$ relative accuracy.
%
%
Because the {two-photon-exchange} contribution can be expressed as the sum of a
finite-size (static, elastic) part proportional to the Zemach radius
($R_Z$) and a polarizability part (dynamic, virtual excitation), its
determination can be used to extract separately the two parts: the Zemach radius
when the polarizability contribution is assumed from theory~\cite{Peset2014, Hagelstein2016, Hagelstein2018, Faustov2002, Cherednikova2002, Peset2017, Tomalak2019, Alarcon2020, Carlson2011}, and the polarizability contribution when taking $R_Z$ from electron-proton scattering or hydrogen {spectroscopy}~\cite{Friar2005, Volotka2005, Distler2011, lin2021}.

  \begin{figure}[t]
  \centering
\includegraphics[width=.65\textwidth]{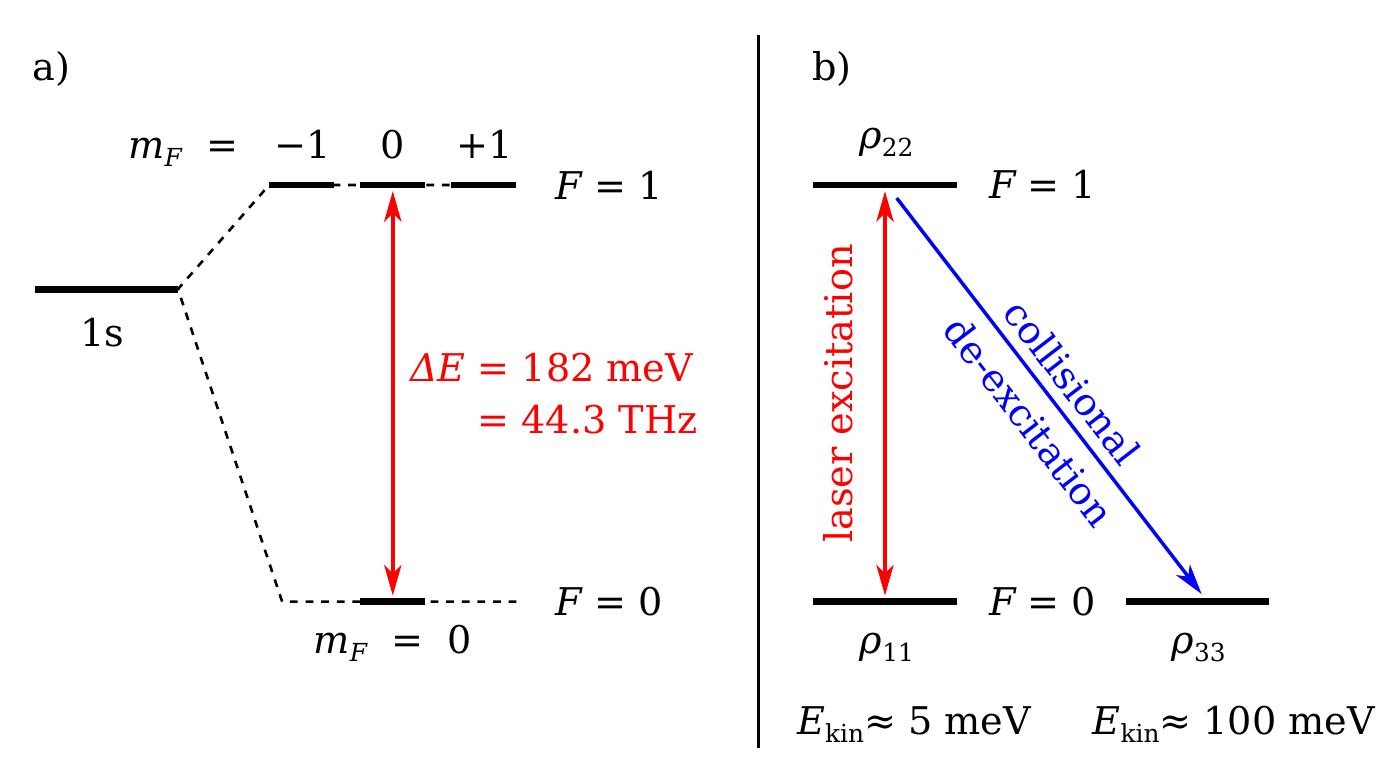}
\caption{(a) Hyperfine structure of the 1s-state in $\mu$p divided into the
  triplet ($F=1$) and the singlet ($F=0$) states depending on the
  total angular momentum of the muon-proton system.  (b)
  The three-level system used in the Bloch equations to model the laser
  excitation followed by collisional deexcitation with an increase of
  kinetic energy ($E_\text{kin}$).  Initially all $\mu$p atoms are thermalized (average of $E_\text{kin}\approx5$~meV) to the
  singlet state with population $\rho_{11}$. The laser pulse drives
  the HFS transition, exciting the $\mu$p atoms into the triplet state
  with population $\rho_{22}$. 
  An inelastic collision then deexcites
  the triplet state back to the singlet state converting the
  transition energy into kinetic energy. This singlet state with
  additional kinetic energy is the third level in the optical Bloch
  equations with population $\rho_{33}$ {and} $E_\text{kin}\approx100$~meV.
}
\label{fig:scheHFS}
\end{figure}

In this paper, we calculate the laser transition probability between
singlet and triplet sublevels of the ground state hyperfine-splitting
in $\mu$p (see Figure~\ref{fig:scheHFS}), accounting for the actual
detection scheme used in the experiments and considering
collisional and Doppler effects.
This transition probability is one of the key quantities needed  to
evaluate the feasibility of the CREMA hyperfine-splitting experiment, and to
define the requirements for the experimental setup.

Two other collaborations, the FAMU collaboration at RAL
(UK)~\cite{Adamczak2001, Dupays2003, Bakalov2015a,
  Bakalov2015b} and another one at J-PARC (Japan)~\cite{Sato2015,
  Saito2012, Kanda2018},  are also aiming at a precision measurement of the HFS
  in $\mu$p.
For all three experimental collaborations, the main challenge is posed by
the small laser-induced transition probability
resulting from the small matrix element (magnetic dipole-type transition, $M1$), in conjunction with
a transition wavelength at 6.8~$\mu$m where no adequate (sufficiently
powerful) laser technologies are available, and the large volume where
$\mu$p atoms are formed.

Prior to this work there were three other publications addressing the
laser-induced excitation probability between the hyperfine states: the
FAMU collaboration performed the calculation assuming the Fermi-golden
rule with Doppler convolution  {while neglecting  collisional effects as well as laser bandwidth}~\cite{Adamczak2012}.
This calculation was revised in Ref.~\cite{Bakalov2018} that corrected
a mistake in the matrix element.
The J-PARC collaboration calculated the laser-induced transition probability using the optical Bloch {equations \cite{MohanDas2020}, also omitting  collisional effects given their  low target density~\cite{Kanda2018}}. Laser bandwidth was also omitted in this calculation. 
In contrast, decoherence effects due to collisions and laser bandwidth were considered in this work, as they reduce the transition probability by almost a factor of two at the optimal experimental conditions. Therefore, collisional rates between $\mu$p and H$_2$ were calculated in this work for the CREMA experimental conditions following the theory of Ref.~\cite{Adamczak2006}.

The paper is  organized as follows: in Sec.~\ref{sec:experiment}
we summarize the CREMA experimental scheme needed to understand the
experimental conditions in which the laser excitation takes place.
 Section~\ref{sec:theory} introduces our theoretical framework based
 on the optical Bloch equations. We also summarize there the
 collisional effects and the transition matrix elements.
 In Sec.~\ref{sec:limits}, starting from the Bloch equations we derive
 well-known analytical expressions valid only in certain
 regimes which serve to better understand the dynamics during the
 laser excitation and the numerical results of Sec.~\ref{sec:numerical}.
The latter section presents the results obtained {by} integrating
numerically the Bloch equations at various experimental conditions
and discusses in detail the impact of collisions, Doppler-effects, and
laser parameters.

\section{The experimental scheme}
\label{sec:experiment}

In this section we briefly present the experimental scheme for the measurement of the HFS 
pursued by the CREMA collaboration, to precisely define the goal and framework
of the calculation.
\begin{figure}[bt]
  \centering
          \includegraphics[width=0.65\textwidth]{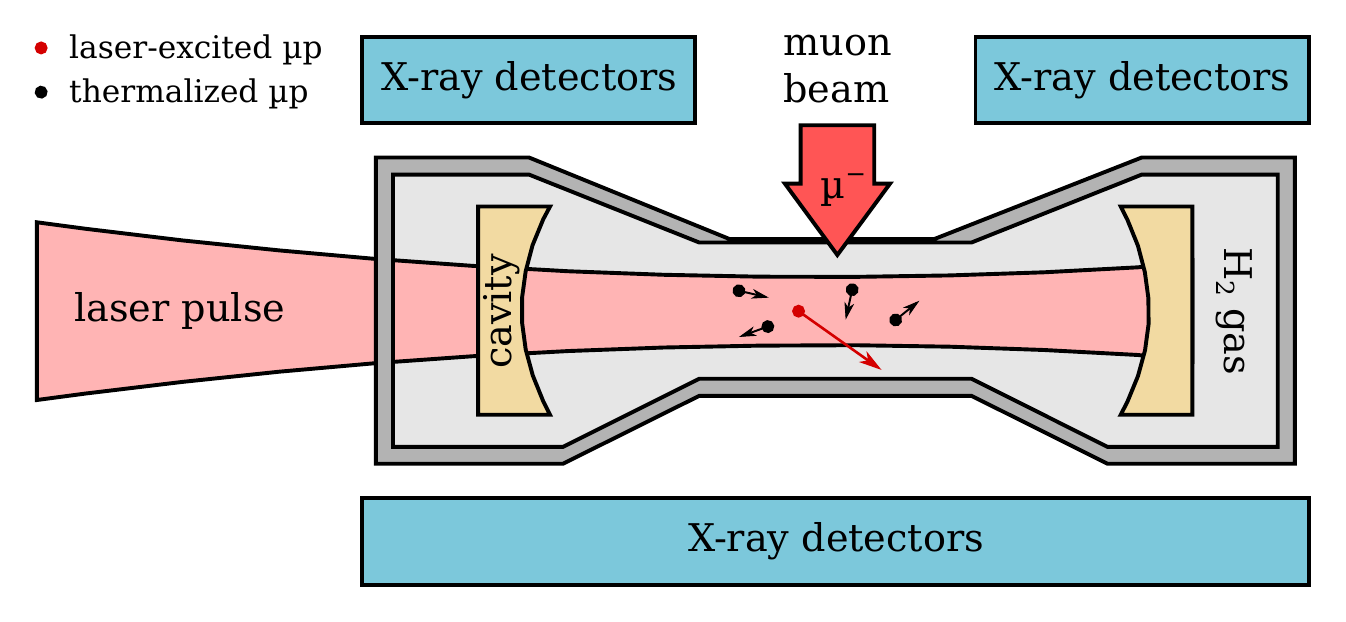}
\vspace{-4mm} \caption{(Color online) Schematic (not to scale) of the setup for
  measuring the HFS in $\mu$p. A negative muon beam is stopped in a
  H$_2$ gas target at cryogenic temperatures, pressures of about 1 bar,
  and a thickness of about 1 mm. The formed $\mu$p atoms are excited by
  the laser pulse whose intensity is enhanced in the multi-pass
  cavity. A successful laser excitation of the $\mu$p atoms followed
  by a collisional deexcitation leads to a $\mu$p atom with extra
  kinetic energy. With this extra kinetic energy, the $\mu$p atom
  diffuses to the target walls where x-rays are produced. }
        \label{fig:target-region}
\end{figure}
A low-energy muon beam of about 11~MeV/c momentum is stopped in a
hydrogen gas target ($\sim 1$~mm thickness, a pressure ranging from $0.5$ to 1~bar,  and
at a temperature ranging from 20 to 50~K), wherein a $\mu$p atom is formed in a 
highly excited state.
The formed muonic atom quickly deexcites to the singlet ($F=0$) state of
the ground state (see Figure~\ref{fig:scheHFS}) while thermalizing
within about 1~$\mu$s to the hydrogen gas temperature.
When thermalized, the $\mu$p {atom}  is illuminated by a laser pulse at a wavelength of 6.8~$\mu$m (equivalent
to a frequency of 44~THz, or {0.18~eV} energy) to drive the hyperfine
transitions.
To enhance the laser transition probability, the laser pulse with a duration of
about 20~ns  and 5~mJ of energy is coupled through a 0.5~mm wide slit
into a multi-pass cavity.
The multiple reflections occurring in this cavity, whose
average number depends mainly on the mirror reflectivity and the
losses at the coupling slit, enhance the laser fluence at the position of the $\mu$p atom and increase the effective pulse length.
On-resonance laser light thus excites the muonic atom from the 
singlet  $F=0$ to the triplet $F=1$  sublevels.
A subsequent inelastic {collision} with a hydrogen molecule deexcites
the $\mu$p atom from the triplet back to the singlet sublevels.
In this process the HFS transition energy of 0.18~eV is converted
into kinetic energy: on average 0.1~eV of kinetic energy is imparted to {the} $\mu$p.
This kinetic energy, which is much larger than the thermal energy, causes the $\mu$p atom
to diffuse away from the {laser-illuminated} volume,  
reaching one of the {gold-coated} target walls in a time window between 100 and 300~ns after the laser excitation.
%
%
%
When the $\mu$p atom reaches {the} wall, the muon is
transferred to a gold atom forming muonic gold ($\mu$Au$^*$) in highly excited
states.
The various x-rays of MeV energy  produced in the subsequent deexcitation 
of $\mu$Au$^*$ are used as signature of a successful laser excitation,
so that the HFS resonance can be determined by counting the number of $\mu$Au
{x-rays} as a function of the laser frequency.

{In this paper, we present the calculation} of the probability that a $\mu$p
atom initially in the singlet state and thermalized at the {temperature of} the hydrogen
gas  will undergo the above described sequence of laser
excitation and collisional deexcitation, acquiring the extra {$\sim 0.1$~eV} of
kinetic energy needed to provide the {observable signal.}

\section{Theoretical framework}
\label{sec:theory}

\subsection{Bloch equations}
The laser excitation and the population dynamics between hyperfine states is
investigated in the framework of the density-matrix formalism, using optical
Bloch equations expressed as~\cite{Loudon2000a},
%
%
{
\begin{eqnarray}
\frac{d \rho_{11}}{dt}(t) &= &  -\mbox{Im}\left(  \Omega \rho_{12} e^{i \Delta t}  \right) + \Gamma_\mathrm{sp}\rho_{22} ~, \label{eq:eqmotim_11}\\
\frac{d \rho_{22}}{dt}(t) &=& \phantom{-}\mbox{Im}\left(  \Omega \rho_{12} e^{i \Delta t}  \right) -  (\Gamma_i+ \Gamma_\mathrm{sp})\rho_{22}~,  \label{eq:eqmotim_22} \\
\frac{d \rho_{12}}{dt}(t) &=& \frac{i \Omega^*}{2} (\rho_{11}-\rho_{22} ) e^{-i  \Delta t} -   \frac{\Gamma_c}{2} \rho_{12} ~, \label{eq:eqmotim_12} \\
\frac{d \rho_{33}}{dt} (t)&=&     \Gamma_i \rho_{22} ~,  \label{eq:eqmotim_33}
%
\end{eqnarray}}
where the detuning $\Delta=\omega_r - \omega$ is the departure of the
laser {angular} frequency $\omega$ from the atomic resonance {angular}  frequency
$\omega_r$, $\Omega$ is the Rabi frequency, $\Gamma_i$ is the triplet
deexcitation rate caused by the inelastic collisions with the H$_2$ gas
leading the $\mu$p atoms to gain  extra kinetic
energy, $\Gamma_\mathrm{sp}$ is the spontaneous radiative decay rate
from the triplet state ({with negligible change} of the kinetic
energy), and $\Gamma_c$ is the  decay rate of the coherence, referred as ''decoherence rate''. All deexcitation and decoherence sources in equations~\eqref{eq:eqmotim_11}-\eqref{eq:eqmotim_33} are in units of Hertz.
Within our experimental conditions, $\Gamma_\mathrm{sp}\ll \Gamma_i$ so
that throughout this paper $\Gamma_\mathrm{sp}$ can be neglected. Note that in equations~\eqref{eq:eqmotim_11}-\eqref{eq:eqmotim_33} we also neglect the muon decay, since  at our experimental {conditions the muon decay rate  is much smaller than the decoherence rate. The overall decay of all populations due to muon decay ($\tau_\mu=2.2$~$\mu$s) can be treated separately simply by multiplying our results by $\exp{(-t/\tau_\mu)}$.}

The diagonal density terms $\rho_{11}$ and $\rho_{22}$ are the
populations of the thermalized singlet  and (excited) triplet states, respectively, while $\rho_{12}$ is the coherence between these two states.

The third level with population $\rho_{33}$ is used to track the
fraction of $\rho_{22}$ that undergoes a collisionally-induced deexcitation
from the triplet to the singlet states, with consequent increase of
the kinetic energy.
%
%
This third level is modeled as a dark state decoupled from the laser 
%
%
because the additional 0.1~eV {of} kinetic energy (corresponding to 
$\sim$1000~K), acquired by the $\mu$p in the inelastic collision, gives
rise to such a large Doppler shift that the atomic resonance becomes
detuned from the exciting laser frequency.
Actually, within several hundred nanoseconds, the $\mu$p atoms {can} thermalize
again through elastic $\mu$p--H$_2$ collisions. {However, $\mu$p atoms with the extra kinetic energy are very likely to exit the laser-excitation volume within tens of nanoseconds.
}
Hence, multiple cycles of laser-excitation followed by
collisional-quenching are negligible at our experimental conditions,
and thus modeling the third level as a dark state is appropriate.

%
As initial conditions for the time $t=0$, we assume $\rho_{11}=1$ and
$\rho_{22}=\rho_{33}=0$ since in the experiment we choose the arrival
time of the laser pulse so that all $\mu$p atoms are deexcited to the
singlet state and thermalized to the hydrogen gas temperature.
As a consequence of $\rho_{22}=0$, for time $t=0$ the coherence $\rho_{12}=0$ is also zero.
In this paper, we assume that the laser excitation occurs from time
$t=0$ until $t=\tau$ at constant laser intensity $\mathcal{I}$,
resulting in a laser fluence of $\mathcal{F}=\mathcal{I} \tau$
[J/cm$^2$].
The impact of the laser fluence $\mathcal{F}$ and exposure time  $\tau$
on the population dynamics is evaluated throughout this paper.
Exposure times  of 10~ns and 100~ns are considered.
Note that the actual time (and spatial) distribution of the laser
intensity within the multi-pass cavity is  complicated by the superpositions of the folded laser beam spreading in the cavity,  however at late times it follows approximatively an exponential function with a
lifetime ranging from 50~ns to 150~ns depending on the performance and
geometry of the cavity.
%

%
  
In this work, the temporal evolution of the populations in the three levels is computed
by numerically integrating the optical Bloch equations  {for various target conditions and laser performances. When combined with simulations of the diffusion process, muon beam and detection system, this allows  optimization of the experimental setup to maximize  statistical significance (signal/$\sqrt{\text{background}}$).  
  }
The parameters $\Gamma_i$ and $\Gamma_c$ allow a phenomenological
inclusion in the Bloch equations of the laser
bandwidth and of the $\mu$p--H$_2$ elastic and inelastic collisions.
The Doppler effect is accounted for in a second 
step by folding the results from the optical Bloch equation with a
Doppler (Gaussian) profile for the (initial) thermalized $\mu$p atoms.

\subsection{Collisional rates}
\label{sec:collisions} 

The decoherence rate is given by 
\begin{equation}
  \Gamma_c={2\pi\Delta_l}+\Gamma_e^{F=0}+\Gamma_e^{F=1}+\Gamma_i+\Gamma_\mathrm{sp}~, 
  \label{eq:gamma_c}
\end{equation}
where {$\Delta_l$ is the laser (FWHM) bandwidth, which contains a $2\pi$ factor  to convert a bandwidth into a ''rate'' (see \cite{Loudon2000a})}. 
%
 $\Gamma_i$ is the inelastic $\mu$p--H$_2$ collision rate for the excited state, $\Gamma_e^\mathrm{F=1}$ is the elastic
collision rate for the triplet,  and 
$\Gamma_e^\mathrm{F=0}$ is the elastic collision rate for the singlet state: 
\begin{eqnarray}
\Gamma_i:  & \quad \mu \mathrm{p}^\mathrm{F=1} + \mathrm{H}_2 \rightarrow \mu
\mathrm{p}^\mathrm{F=0}+\mathrm{H}_2^*\\
\Gamma_e^\mathrm{F=1}:  & \quad  \mu \mathrm{p}^\mathrm{F=1} + \mathrm{H}_2 \rightarrow \mu
\mathrm{p}^\mathrm{F=1}+\mathrm{H}_2^*\\
\Gamma_e^\mathrm{F=0}:   & \quad \mu \mathrm{p}^\mathrm{F=0} + \mathrm{H}_2 \rightarrow \mu
\mathrm{p}^\mathrm{F=0}+\mathrm{H}_2^* \, .
\end{eqnarray}
Here the $^*$ indicates possible rotational
excitations of the hydrogen molecule after the collision. Note that these collisional cross-sections do not depend  on the magnetic sub-states  of  the $\mu$p atom \cite{Bracci1989, Bracci1990}. 
%
%

All the collisional rates {($\Gamma_i$, $\Gamma_e^{F=1}$ and $\Gamma_e^{F=0}$)} were obtained using
\begin{equation}
\Gamma= \overline{\upsilon_r \sigma(\upsilon_r)} \rho_{\mbox{\tiny H}_2} \; , 
\end{equation}
where $\upsilon_{r}$ is the $\mu$p--H$_2$ relative velocity
and $\rho_{\mbox{\tiny H}_2}$ is the number density of H$_2$ molecules, which was calculated assuming that H$_2$  behaves as an ideal gas.
%
The term $\overline{\upsilon_{r} \sigma(\upsilon_r)}$ represents the folding of the
relative $\mu$p--H$_2$ velocity ($\upsilon_r$) distribution for thermalized
$\mu$p atoms and H$_2$ molecules with the velocity-dependent 
cross sections $\sigma (\upsilon_r)$ for the considered scattering process
~\cite{Adamczak2006}.

These {rates} were calculated following Ref.~\cite{Adamczak2006}, which uses
the velocity- and spin-dependent cross sections between $\mu$p
atoms and hydrogen molecules summed over the final 
rotational states of the H$_2$ molecules and averaged
over the distribution of the initial rotational states.
The spin alignment of the two nuclei of the molecule is taken into consideration when
computing the rotational states and their initial populations.
%
{Two distributions for the initial rotational states were considered: statistical and Boltzmann distributions. At room temperature, the Boltzmann distribution is similar to the statistical distribution given by 75\%  ortho-hydrogen (odd rotational number $K$) and 25\%  para-hydrogen (even $K$). For low temperatures and at equilibrium, the  Boltzmann distribution differs considerably from the statistical distribution, e.g. at 22~K,  basically all molecules  have $K=0$. 
However, if the gas is quickly cooled down, the ratio ortho- to para-hydrogen existing at high temperature is retained so that 75\% of the molecules have $K=1$ and 25\% have $K=0$.
The Boltzmann distribution is only reached after a long time because the conversion  from $K=1$ to $K=0$ rotational states is a very slow process having a rate of about 2\% per week at the normal temperature
and pressure  \cite{Silvera1980}.}
%
The actual rotational distribution in the experimental conditions is {therefore} bounded between these two extreme cases.
%
%
For this reason, we give in Table~\ref{tab:2} values of the collision
rates for statistical and Boltzmann distributions. 


\begin{table}[t]
 \caption{$\mu$p--H$_2$ collision rates in MHz for various H$_2$ target
   pressures ($p$) and temperatures ($T$). Two distributions of
   ortho- and para-H$_2$ are considered: statistical and Boltzmann.  }
\centering
\begin{tabular} {lccccccccccc} 
\\[-2.0ex]  \hline  \hline  \\[-2.0ex]									
\multicolumn{1}{l}{$p=0.5$ bar} \\													
	&	\multicolumn{2}{c}{$T=22$~K}			&&	\multicolumn{2}{c}{$T=30$~K}		&&	\multicolumn{2}{c}{$T=50$~K}				\\
													
	&	Stat.	&	Boltz.	&\;&	Stat.	&	Boltz.	&\;&	Stat.	&	Boltz.	\\
$\Gamma_e^{F=0}$	&	20	&	20	&&	15	&	15	&&	9	&	9	\\
$\Gamma_e^{F=1}$	&	52	&	29	&&	41	&	24	&&	28	&	18	\\
$\Gamma_i$	&	82	&	93	&&	59	&	66	&&	34	&	37	\\
		\hline \\[-2.0ex]											
		\multicolumn{1}{l}{$p=1$ bar} \\											
$\Gamma_e^{F=0}$	&	40	&	39	&&	30	&	30	&&	19	&	19	\\
$\Gamma_e^{F=1}$	&	104	&	59	&&	83	&	47	&&	55	&	37	\\
$\Gamma_i$	&	164	&	187	&&	118	&	133	&&	68	&	74	\\
\hline \\[-2.0ex]													
		\multicolumn{1}{l}{$p=2$ bar} \\											
$\Gamma_e^{F=0}$	&	79	&	79	&&	61	&	61	&&	38	&	37	\\
$\Gamma_e^{F=1}$	&	208	&	118	&&	165	&	94	&&	110	&	74	\\
$\Gamma_i$	&	328	&	374	&&	235	&	265	&&	137	&	148	\\
\\[-2.0ex]  \hline  \hline  \\[-2.0ex]
\end{tabular}
  \label{tab:2} 
\end{table}

 \begin{table}
\centering
\caption{Values of ${\mathcal{M}}$  and  $\Omega/\sqrt{\mathcal{I}}$   for Lamb shift and HFS transitions in $\mu$p and $\mu^3$He$^+$. The analytical expressions for the $2s-2p$ matrix elements  agree with \cite{Schmidt2018a}. $a_\mu$ is the muonic Bohr radius.   
  \label{tab:m_matri} }
\begin{tabular} {llll} 
\\[-2.0ex]  \hline \hline  \\[-2.0ex]	 
Atom & Transition & ${\mathcal{M}}$ [m]  & $\frac{\Omega}{\sqrt{\mathcal{I}}}$ [$\mbox{m}/\sqrt{\mbox{Js}}$] \\
\\[-2.0ex]  \hline  \\[-2.0ex]	 
$\mu$p &$2s^{F=1}\rightarrow 2p_{3/2}^{F=2}$ &   $\sqrt{5}a_\mu=6.367\times10^{-13}$   & 2.65 $\times10^{4}$  \\
%
$\mu^3$He$^+$& $2s^{F=1}\rightarrow 2p_{3/2}^{F=2}$ & $\frac{\sqrt{5}}{2} a_\mu=2.969\times10^{-13}$ & 1.24 $\times10^{4}$ \\
%
    $\mu$p &  $1s^{F=0} \rightarrow 1s^{ F=1}$  & 
$\frac{ \hbar}{4 m_\mu c } \left( g_\mu + \frac{m_\mu}{m_p} g_p \right)$  &  5.12 $\times10^{1}$ ~$^a$\\ 
  %
  &  & $=1.228\times10^{-15}$  & \\ 
  %
$\mu^3$He$^+$ &  $1s^{F=1} \rightarrow 1s^{ F=0}$  & $\frac{ \hbar}{4\sqrt{3} m_\mu c} \left( g_\mu + \frac{m_\mu}{m_{\tiny \mbox{He}}} g_{\tiny \mbox{He}} \right)$ &  2.07 $\times10^{1}$ \\
%
&& $=4.965\times10^{-16}$\\   
\\[-2.0ex]  \hline \hline  \\[-2.0ex]	 
\end{tabular}
\\$^a${ 1.77$\times10^{1}$ $\mbox{m}/\sqrt{\mbox{Js}}$ according  to Ref.~\cite{MohanDas2020}} 
\end{table}

    

\subsection{Rabi  frequencies and matrix elements}
\label{sec:rabi_freq} 
The Rabi frequency $\Omega = \sqrt{\frac{8 \pi \alpha \mathcal{I} }{
    \hbar} } {\mathcal{M}}$ included in the Bloch equations
%
%
%
%
quantifies the coupling strength between the laser with
intensity $\mathcal{I} $, and the atomic transition.
%
Here, $\alpha$ is the fine structure constant and $\hbar$ the reduced
Planck constant.
We evaluated the matrix element  ${\mathcal{M}}$ for the M1 HFS
transitions  of $\mu$p and $\mu^3$He$^+$ in the ground state.
To verify the calculations, we also evaluated  the
matrix elements for the already measured $2s-2p$ transitions in $\mu$p
and $\mu^3$He$^+$, where literature values of saturation fluences are available~\cite{Schmidt2018a}.

Because the magnetic substates of the initial state can be assumed to
be statistically populated, the matrix element is averaged over the
initial magnetic sub-states $m$:
\begin{equation}
{\mathcal{M}}^2=\frac{1}{2F_i +1}\sum_{m, m'}\left| \mathcal{M}^{(m, m')}\right|^2. 
\end{equation}
It is also summed  over all possible final
magnetic sub-states $m'$ since the final magnetic sub-states are not detected (resolved) in our experimental scheme.
The matrix elements $\mathcal{M}^{(m, m')}$ for  the $2s-2p$ (E1-type) and  $1s$-HFS (M1-type) transitions  are given by
\cite{Johnson2007, Adamczak2012}
\begin{eqnarray}
&&\mathcal{M}_{\mbox{E1}}^{(m, m')} =  \left\langle 2p\, {F'} \,m' \left| \bm{r}\cdot \bm{\hat{\varepsilon}} \right| 2s \, {F} \, m \right\rangle,  \nonumber \\
 && \mathcal{M}_{\mbox{M1}}^{(0, m')} =  \label{eq:mar_E1M1}   \\  
&&  \frac{1}{2m_{\mu}c}\left\langle F'=1\, m' \left| (g_{\mu}\bm{S} + g_{p}\frac{m_{\mu}}{m_p} \bm{I} )\cdot\bm{\hat{k}\times \hat{\varepsilon} }
   \right| F=0 \,0 \right\rangle,  \nonumber 
\end{eqnarray}
where $m_\mu$ is the muon mass, $m_p$ the proton mass, $\bm{\hat{k}}$
the laser wavevector, $\bm{\hat{\varepsilon}}$ the laser polarization,
$\bm{S}$ the spin operator of the muon, $\bm{I}$ the spin operator
of the proton, and $g_{\mu}$ and $g_{p}$ the $g$-factor of the muon
($\simeq 2.00$) and proton ($\simeq 5.58$), respectively.
See Appendix~\ref{sec:appen} for more details about the evaluation of these matrix elements.

Table \ref{tab:m_matri} summarizes the numerical values of the (analytical) matrix
elements ${\mathcal{M}}$ and $\Omega/\sqrt{\mathcal{I}}$ for the
HFS and the most intense Lamb shift transitions for  $\mu$p and
$\mu^3$He$^+$.
While the matrix elements for the $2s-2p$ transitions are in agreement
with published values~\cite{Schmidt2018a}, the $\Omega/\sqrt{\mathcal{I}}$ value
obtained for the $\mu$p HFS of $5.12\times 10^{1}
\;\mbox{m}/\sqrt{\mbox{Js}}$ is in disagreement with the value of $1.77\times 10^{1}
\;\mbox{m}/\sqrt{\mbox{Js}}$ published in
Ref.~\cite{MohanDas2020}.
{The  difference} of $\sqrt{\frac{3}{2} }\left( \frac{g_\mu m_p +g_p m_\mu
}{m_p + m_\mu} \right)\approx 2.9$ is traced back to
a miscalculation of the matrix elements in Ref.~\cite{MohanDas2020} {(see
Appendix~\ref{sec:appen}), and by not considering the proton spin-flip.}

\subsection{Doppler broadening }
\label{sec:broade_Dopp}

%
While the collisional effects and the laser bandwidth are already
accounted for in the Bloch equations, and eventually give rise to a
Lorentzian profile (see Sec.~\ref{sec:previuos_cal_Fer}), the Doppler broadening needs to be
treated separately. 
 \begin{table*}[t]
 \caption{Saturation fluences  for laser {bandwidths} of $\Delta_l=10$~MHz ($\mathcal{F}_{\mbox{sat}}^{\Delta_l\rightarrow 10 }$) and $\Delta_l=100$~MHz ($\mathcal{F}_{\mbox{sat}}^{\Delta_l\rightarrow 100 }$)   for three transitions in muonic atoms{, accounting for collisional effects}. For comparison  we also give the saturation fluence neglecting decoherence effects ($\mathcal{F}^{\Gamma_c\rightarrow 0 }_{\mbox{sat}}$).  The resonance frequencies {$\nu_{r}=\omega_{r}/2\pi$}  for the three  transitions have been taken  from Refs.~\cite{Borie2012a,Antognini2013b, Martynenko2008}. The spontaneous decay  rates $\Gamma_{\mbox{sp}}$  of the  $2p$ states were taken from \cite{Milotti1998a} and   \cite{Amaro2015a}, while  for {the ground state}  triplet in $\mu$p it is calculated here.  The decoherence rates $\Gamma_c$ have been obtained using equation~\eqref{eq:gamma_c} with the values from Table~\ref{tab:2}  assuming a Boltzmann distribution between ortho- and para-H$_2$ (for a statistical distribution see Appendix~\ref{sec:Appenstatistic}). {$\sigma_D$ is the Doppler standard deviation  \eqref{eq:Omega}.    } 
   }
      \resizebox{\textwidth}{!}{
\begin{tabular} {lccccccccccc} 
\\[-2.0ex]  \hline  \hline  \\[-2.0ex]
Atom &Transition & $T$   & $p$      & {$\nu_{r}$} & $\Gamma_{\text{sp}}$  & $\Gamma_c^{\Delta_l\rightarrow10}$ & $\Gamma_c^{\Delta_l\rightarrow100}$  & $\sigma_D$ & $\mathcal{F}^{\Gamma_c\rightarrow 0 }_{\mbox{sat}}$ & $\mathcal{F}_{\mbox{sat}}^{\Delta_l\rightarrow 10 }$& $\mathcal{F}_{\mbox{sat}}^{\Delta_l\rightarrow 100 }$ \\
     &                & [K] & [bar]  & [THz]        & [MHz]   & [MHz] & [MHz]  & [MHz] & [J/cm$^2$] & [J/cm$^2$] & [J/cm$^2$] \\
  [+1.0ex]  \hline	\\ [-1.0ex]
$\mu$p        & $2s^{F=1}\rightarrow 2p_{3/2}^{F=2}$  & 300  & 0.001  & 49.9 &  {1.16$\times10^{5}$} &  & {1.16$\times10^{5}$} & {2.48$\times10^{2}$}  &    & & 0.0165 \\  
$\mu^3$He$^+$   & $2s^{F=1}\rightarrow 2p_{3/2}^{F=2}$  & 300  & 0.004  & 379  & {2.00$\times10^{6}$} & &  {2.00$\times10^{6}$}  & {1.13$\times10^{3}$}  &     & &1.304 \\  
  [+1.0ex]  \hline	\\ [-1.0ex]
$\mu$p & $1s^{F=0} \rightarrow 1s^{ F=1}$           &        &       & 44.2 & {1.23$\times10^{-11}$} & &    &      &        \\  
       &                                           &  22   & 0.5   &      &              &   {205}     & {770} & 60   & 23 & {28} & {44} \\ 
       &                                           &  22   & 1      &      &              &   {348}    &{913} & 60   & 23 & {32} &{49} \\ 
       &                                           &  22   & 2     &       &              &   {633}   & {1198} & 60   & 23 & {40} & {58}\\[1mm] 
       &                                           &  30   & 0.5   &      &              &  {168}   & {733} & 70   & 27 & {31} & {47} \\
       &                                           &  30   & 1      &      &              &  {273}   & {838} & 70   & 27 & {34} & {50} \\[1mm]
       &                                           &  50   & 0.5   &      &              &  {128} & {693} & 90   & 35 &  {37} &{53} \\
       &                                           &  50   & 1      &      &              & {192} & {757} & 90   & 35 &  {39} &  {55} \\

 \\[-2.0ex]  \hline  \hline  \\[-2.0ex]
\end{tabular}
}
  \label{tab:flue_E1} 
\end{table*}

We include the Doppler effect by convoluting the population
${\rho}_{33}(\omega)$ as obtained from the Bloch equations with a
Gaussian distribution describing the Doppler profile:
  {
 \begin{equation}
\bar{\rho}_{33}(\omega)=\int_{-\infty}^{\infty} \rho_{33}(\omega')\frac{1}{\sqrt{2\pi}\gamma_D}\exp\left(- \frac{(\omega-\omega')^2}{2\gamma_D^2}\right)d\omega'~,
\label{eq:inhome_Doppler}
\end{equation}
with $\gamma_D$ being  given by
\begin{eqnarray}
 \gamma_ D=\omega_{r}\sqrt{\frac{ k T}{(m_\mu + m_p)c^2}}&\simeq& 7.98\times 10^{7}\sqrt{T}~\text{[rad/s]}, \\
   \sigma_ D= \frac{\gamma_ D}{2\pi} &\simeq& 12.7 \sqrt{T}\, ~\text{[MHz]}, 
 \label{eq:Omega}
\end{eqnarray}}
so that the Doppler standard deviation (MHz) is $\sigma_D$.  $k$ is the Boltzmann constant, $c$ the speed of light, and $T$
the temperature in Kelvin of the thermalized $\mu$p atom in the singlet state. 
%
%


\section{Analytical expressions for two limiting regimes}
\label{sec:limits}



Before integrating numerically the three-level Bloch equations and
investigating how collisional and Doppler effects impact the $\rho_{33}$
population, it is interesting to derive analytical expressions from
our formalism to reproduce {well-known} results valid for low and high
Rabi frequencies, where low and high are relative to the other
frequencies involved in the problem.


\subsection{Fermi-golden-rule regime}
\label{sec:previuos_cal_Fer}
The laser-induced rate of population transfer $R(t,\omega)$ from
$\rho_{11}$ to $\rho_{22}$ is given by the first term of the second
Bloch equation \eqref{eq:eqmotim_22}, 
\begin{equation}
R(t,\omega)=\mbox{Im}\left(  \Omega \rho_{12} e^{i \Delta t}  \right)~.
\label{eq:R_gen}
\end{equation}
In the limit of low transition probability ($\Omega \ll \Gamma_c$),
and sufficiently long time ($t\gg 1/\Gamma_c$) to obtain  stable $\rho_{22}$ population,
this rate converts  to the Lorentzian form of the Fermi-golden
rule~\cite{Loudon2000a},
\begin{equation}
R_\text{F}^\text{(L)} ( \omega)= \frac{ |\Omega|^2 }{4} \frac{\Gamma_c}{\left(\Delta^2 +\left(\frac{\Gamma_c}{2}\right)^2  +  \frac{|\Omega|^2 \Gamma_c}{2 \Gamma_i} \right)}   ~.
\label{eq:rate_Lorent}
\end{equation}
This result is obtained using the analytical solutions of  $\rho_{22}$ and  $\rho_{12}$ of Refs.~\cite{Smith1978,
  Smith1979,Smith1980}.
Note that the Fermi-golden-rule approach does not contain  population {dynamics}; it is a static approach.

The Doppler effect can be included by performing a convolution similar to
 equation~(\ref{eq:inhome_Doppler})
 {\begin{equation}
 \bar{R}_\text{F}^\text{(L)}(\omega)=\int_{-\infty}^{\infty} R_\text{F}^\text{(L)} ( \omega') \frac{1}{\sqrt{2\pi}\gamma_D}\exp\left(- \frac{(\omega-\omega')^2}{2\gamma_D^2}\right)d\omega'~,
 \label{eq:inhome_Doppler-1}
\end{equation}}
resulting in a combined laser-excitation and collisional deexcitation probability  of
 \begin{equation}
 \bar{\rho}_{33}(\omega)  \approx   \hat{\rho}_{33}(\omega) =\int_0^\tau \bar{R}_\text{F}^\text{(L)} (\omega)d t=\frac{\mathcal{F}}{\mathcal{F}_{\small \mbox{sat}}(\omega)}~,
\label{eq:P_exc_sim}
\end{equation}
 where $\mathcal{F}_{\small \mbox{sat}}$ is the saturation fluence and
 $\bar{\rho}_{33}(\omega)$ is the obtained Voigt profile that accounts for Lorentzian (homogeneous) and Gaussian (inhomogeneous) broadenings.
The hat-symbol on the third level population $\hat{\rho}_{33}$ is used to
 denote that it was computed in the
 Fermi-golden-rule approximation and including Doppler effects.

Since the Voigt function at resonance has an analytical solution, at resonance
we can find an analytical expression for
$\mathcal{F}_{\mbox{sat}}(\omega_r)$:  
  {\begin{eqnarray}
\mathcal{F}_{\mbox{sat}}(\omega_r)=\frac{\hbar \gamma_D^2 }{  \pi^{3/2} \alpha  \Gamma_c \mathcal{M}^2} \frac{ \varpi e^{-\varpi^2} }{ \mbox{Erfc}(\varpi)}~,
\label{eq:fluenc_doopl}
\end{eqnarray}}
where   {$\varpi=\frac{\sqrt{2}\Gamma_c}{4 \gamma_D}$} and Erfc is the complementary error function.
%
%
For $\Gamma_c \ll \sigma_D$ the saturation fluence becomes
 { \begin{eqnarray}
\mathcal{F}^{\Gamma_c\rightarrow 0}_{\mbox{sat}}(\omega_r)=\frac{\hbar \sigma_D }{  (2\pi)^{3/2} \alpha  \mathcal{M}^2}\approx \frac{\sqrt{T}}{0.2}\;~{[\text{J/cm}^2]}.
\label{eq:fluenc_justDopp}
\end{eqnarray}}
 %
 %
 In this  {limit},  the laser transition probability becomes
 \begin{eqnarray}
 \hat{\rho}_{33}^{\tiny{ \Gamma_c\rightarrow 0}}  \approx \frac{0.2}{\sqrt{T}}\mathcal{F} \; ,
 \label{eq:FAMU-probability}
 \end{eqnarray}
 {where $T$ and $\mathcal{F} $ are values of temperature and fluence in units of K and J/cm$^2$, respectively. }
 This result is in agreement with the excitation probability given
 in Ref.~\cite{Bakalov2018}.
 %

 Table~\ref{tab:flue_E1} summarizes the saturation fluences
 $\mathcal{F}_{\small \mbox{sat}}$ calculated from
 equation~\eqref{eq:fluenc_doopl}, for Lamb shift transitions in $\mu$p and
 $\mu$He$^+$, as well as for the HFS transition in $\mu$p for possible
 experimental conditions.
   Obtained values of $\mathcal{F}_{\small \mbox{sat}}$ for Lamb shift transitions are in agreement with Ref.~\cite{Schmidt2018a}. {Four} orders of magnitude separate the
 $\mathcal{F}_{\small \mbox{sat}}$ for the Lamb shift and the HFS
 transitions in $\mu$p, emphasizing the laser technology leap needed to accomplish  the hyperfine experiment.
 Values of $\mathcal{F}^{\Gamma_c\rightarrow
   0}_{\mbox{sat}}$ given by equation~\eqref{eq:fluenc_justDopp}
 are also shown together with $\mathcal{F}_{\mbox{sat}}^{\Delta_l\rightarrow 10 }$ and $\mathcal{F}_{\mbox{sat}}^{\Delta_l\rightarrow 100 }$, which are the saturation  fluences for $\Delta_l=10$~MHz  and 100~MHz, respectively. The comparison of  $\mathcal{F}_{\mbox{sat}}^{\Delta_l\rightarrow 10 }$ to $\mathcal{F}_{\mbox{sat}}^{\Delta_l\rightarrow 100 }$ highlights the impact of the laser bandwidth, while comparison of  $\mathcal{F}_{\mbox{sat}}^{\Delta_l\rightarrow 10 }$ to $\mathcal{F}^{\Gamma_c\rightarrow 0}_{\mbox{sat}}$ allows to appreciate the impact of the collisional effects. 
%

{Table~\ref{tab:flue_E1_stat} in Appendix \ref{sec:Appenstatistic} shows that {within} our experimental conditions, the decoherence rates and the saturation fluences are only marginally affected  by the assumed ortho- to para-H$_2$ distributions. Therefore, throughout this study, we assumed a Boltzmann distribution.  
 }

\subsection{Rabi-oscillation regime: Two-level Bloch equations}
\label{sec:two_level_B}

For $\Gamma_c < \Omega$, the approximation of equation~(\ref{eq:P_exc_sim}) is no longer valid.
In this limit, the $\rho_{22}$ population saturates and oscillates  according to
\begin{equation}
\rho_{22}= \frac{ |\Omega|^2}{ |\Omega|^2  + \Delta^2} \sin^2\left(\frac{\tau}{2} \sqrt{|\Omega|^2  + \Delta^2}\right)\; ,
\label{eq:rate_osci}
\end{equation}
which on resonance simplifies to 
 \begin{eqnarray}
{\rho}_{22} 	= \sin^2 \left(\sqrt{\frac{8 \pi \alpha \mathcal{F} \tau }{ \hbar}  } \mathcal{M} \right)\; .
\label{eq:R_Rabi_simp}
 \end{eqnarray}
These are the well-known analytical solutions of the two-level
Bloch equations valid in the Rabi-oscillation regime when broadening
 sources can be neglected~\cite{Loudon2000a}, i.e., when the Rabi
 frequency {dominates} over the broadening rates $\Omega \gg
 \Gamma_c$ and $\Omega \gg \sigma_D$.


Reference~\cite{MohanDas2020} calculates the transition probability
between hyperfine states in $\mu$p using equation~(\ref{eq:rate_osci}),
hence {without} decoherence (collisional or laser bandwidth) effects.
%
Doppler broadening was included in a second step in their calculations
as a not-well-specified numerical average.


%
\section{Results and discussion}
\label{sec:numerical}

%
%
%
%
%

\subsection{$\bar{\rho}_{33}$ at the experimental conditions}
\label{sec:calcu_experi}

{Simulating the time-evolution of state populations using Bloch equations has improved laser spectroscopy of molecules  \cite{Rezai2019}, muonium \cite{Nishimura2021} and highly charged ions \cite{Micke2020}. Similarly, the findings of population dynamics obtained from Eq.~\eqref{eq:eqmotim_11}-\eqref{eq:eqmotim_33} can be used to optimize the target conditions ($T$ and $p$) of the HFS experiment.}



We  {present here} the $\bar{\rho}_{33}$ populations
for laser parameters and hydrogen target conditions in the {current} region of
interest for the CREMA HFS experiment.
\begin{figure}[tb]
\centering
\includegraphics[width=1.0\textwidth]{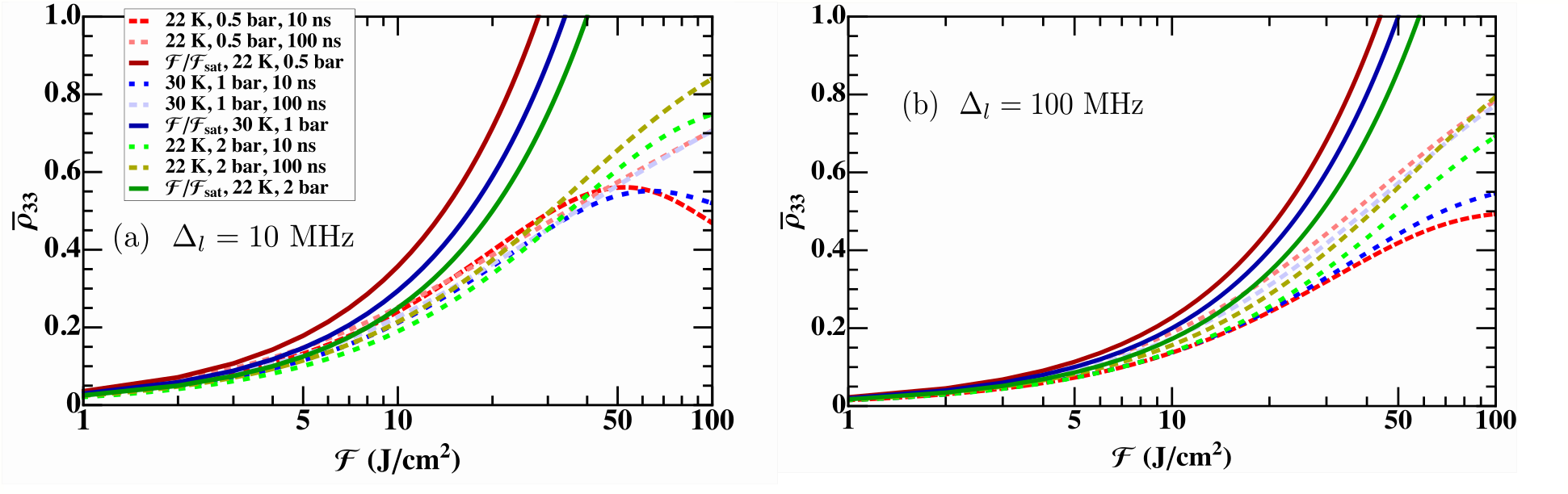}
\caption{ (Color online) $\bar{\rho}_{33}$ population versus laser
  fluence $\mathcal{F}$ for various H$_2$ target conditions,  {two laser bandwidths ($\Delta_l=10$~MHz and $\Delta_l=100$~MHz)}, and two  exposure times
  ($\tau=10$~ns and $\tau=100$~ns) in the
  range of interest for the CREMA HFS experiment.   We use the
  collision rates of Table~\ref{tab:2} assuming Boltzmann
  distribution between ortho- and para-H$_2$. The solid lines represent the
  results from the analytical expression of
  equation~\eqref{eq:P_exc_sim} that neglects population dynamics. The saturation fluences $\mathcal{F}_\text{sat}$   {are taken from  Table~\ref{tab:flue_E1}.}}
\label{fig:compare-analytical-numerical-Fermi-regime}
\end{figure}
These populations obtained from the numerical integration of the
Bloch equations are shown by the dashed curves in
Figure~\ref{fig:compare-analytical-numerical-Fermi-regime} and compared
with the approximations of equation~\eqref{eq:P_exc_sim}, shown as solid lines. 
 {Already for  fluences larger than 5 J/cm$^2$} the full-numerical calculations {need} to be used for accurate predictions. At the prospected  optimal conditions of $T=22$~K and $p=0.5$~bar (based on preliminary simulations of the diffusion of the $\mu$p atom in the H$_2$ gas target),  {for $\Delta_l=10$~MHz}, the  analytical prediction overestimate the transition probability by  about  {30\% and 46\%  for fluences of 10~J/cm$^2$ and 20~J/cm$^2$, respectively}. For $\Delta_l=100$~MHz, this difference is reduced by  {17\% and 26\%} for the same fluences, respectively.  {Note that fluences  up to 20~J/cm$^2$} are  achievable  locally in our multi-pass cavity.  {The importance of the dynamics is also highlighted by the small but non-negligible dependence of $\bar{\rho}_{33}$ on the exposure time $\tau$.}


\subsection{Resonance profile  {at the experimental conditions}}
\label{sec:ress_profile}

The HFS resonance has to be searched for in a region of about 40~GHz ($\pm
3\sigma$) given by the uncertainty of the theoretical predictions
arising mainly from the uncertainty of the two-photon exchange
contribution~\cite{Peset2017, Tomalak2017, Tomalak2019, Hagelstein2016, Karshenboim2014, Alarcon2020, Volotka2005, Dupays2003, Peset2014, Hagelstein2018, Carlson2011}.
The resonance is scanned by counting the number of   {\mbox{x-rays}} for a few
hours at a given laser frequency.
Then the laser frequency is shifted by a fixed amount and the x-ray
counting is performed anew.
This is repeated until a statistically significant deviation of the
number of x-ray counts from the background level has been observed.
Given the large search range, the long time {(few hours)} per frequency point
{necessary} to observe a statistically significant deviation above
background, and the limited access at the PSI accelerator, it is
important to optimize the frequency step used to search for the
resonance.
For this reason, the linewidth of the targeted resonance has to be known precisely.
 \begin{figure}[t]
  \centering
\includegraphics[clip=true,width=1.0\textwidth]{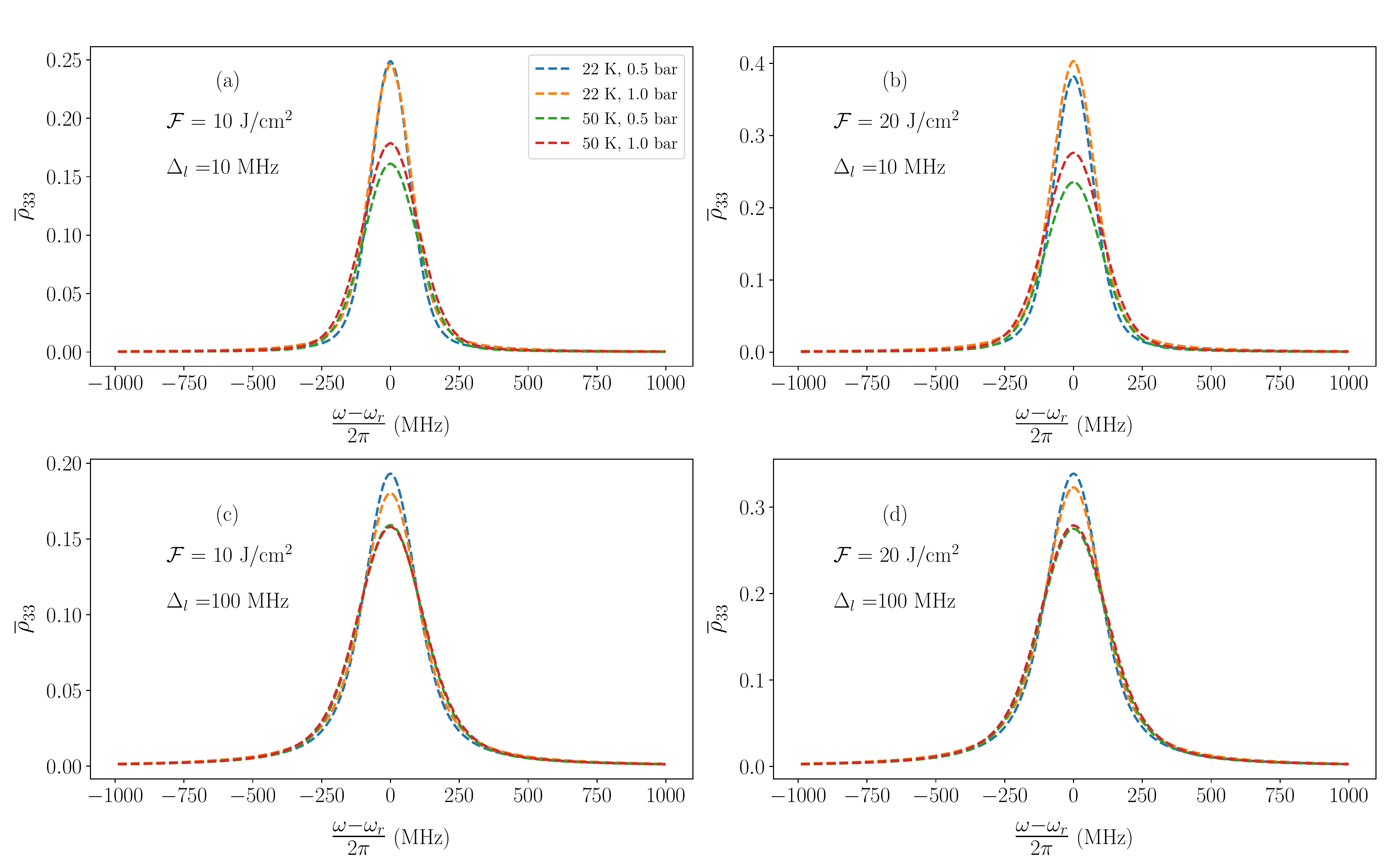} 
\caption{(Color online)  $\bar{\rho}_{33}$ versus the laser angular frequency $\omega$ (line-shape)
  calculated by numerically integrating the Bloch equation for $\tau=100$~ns, at various target conditions
  assuming Boltzmann distributions between ortho- and para-H$_2$:  {{(a) $\mathcal{F}=10$~J/cm$^2$ and $\Delta_l=10$~MHz; (b) $\mathcal{F}=20$~J/cm$^2$ and $\Delta_l=10$~MHz; (c) $\mathcal{F}=10$~J/cm$^2$ and $\Delta_l=100$~MHz;  (d) $\mathcal{F}=20$~J/cm$^2$ and $\Delta_l=100$~MHz}. Values of $\bar{\rho}_{33}$ on resonance and FWHM linewidth can be found in Table \ref{tab:widths}.}
   }
\label{fig:line_profile}
\end{figure}

Figure~\ref{fig:line_profile}  shows some line profiles
obtained from numerical integration at 
conditions relevant for the CREMA experiment.
The figure clearly shows that the reduction of the on-resonance
excitation probability due to collisional effects is correlated with a
broadening of the line-shape.

For fluences $\mathcal{F}\lesssim 20$~J/cm$^2$, the Rabi frequency is
sufficiently small so that this power broadening can be neglected.
In this regime the line shape is thus a Lorentzian profile with a FWHM of approximately  
\cite{Loudon2000a}
\begin{equation}
{\Gamma}_L^2=\Gamma_c^2 + 2|\Omega|^2 \frac{  \Gamma_c}{\Gamma_i}\; .
\label{ref:gamma_L}
\end{equation}
When including the Doppler effect, a Voigt profile is obtained with
a  FWHM-linewidth $\hat{\Gamma}_V$ given approximately by  \cite{Olivero1977}:
\begin{equation}
  \hat{\Gamma}_V=0.53\Gamma_L + \sqrt{0.22\Gamma_L^2 + 5.54   {\gamma_D^2} } \; .
  \label{ref:gamma_V}
\end{equation}
The $\hat{\Gamma}_V$, calculated using the analytical expressions of
Eqs.~\eqref{ref:gamma_L} and ~\eqref{ref:gamma_V} are compared in Table~\ref{tab:widths} to the
FWHM linewidth obtained by numerically
integrating the optical Bloch equations.
%
For completeness, Table~\ref{tab:widths} also summarizes the
$\hat{\rho}_{33}(\omega_r)$ and $\bar{\rho}_{33}(\omega_r)$ as
obtained from the analytical expression of
equation~\eqref{eq:P_exc_sim} and from the full numerical
calculation, respectively. {Here, a Boltzmann distribution between ortho- and para-hydrogen was assumed while the respective values for a statistical distribution are listed in Appendix~\ref{sec:Appenstatistic}. Their comparison  shows a negligible dependence on the type of distribution assumed.} Tables~\ref{tab:widths}, ~\ref{tab:widths_sta} and \ref{tab:widths_time} provide the complete information needed to
quantify the combined probability of laser-excitation followed by collisional
deexcitation for the HFS experiment of the CREMA collaboration. 
 \begin{table*}[t]
   \caption{$\hat{\Gamma}_V$  from the analytical expression of
     equation~\eqref{ref:gamma_V}, FWHM line-width $\Gamma_V$ extracted
     from fitting the line-profile obtained from numerical integration
     of the  Bloch equations for various $\omega$, $\hat{\rho}_{33}(\omega_r)$ from
     the analytical expression of equation~\eqref{eq:P_exc_sim} and
     $\bar{\rho}_{33}(\omega_r)$ from the numerical integration of the
     Bloch equations.  $\tau=100$~ns  {and} Boltzmann distribution between
     ortho- and para-H$_2$ are assumed (for $\tau=10$~ns and a statistical distribution see Appendix~\ref{sec:Appenstatistic}). The symbol “-“ indicates that  $\hat{\rho}_{33}>1$, which  is non-physical.  }
   \resizebox{\textwidth}{!}{
\begin{tabular} {cccccccccccccccccccc} 
\\[-2.0ex]  \hline  \hline  \\[-2.0ex]
&&&& &\hspace{0.5cm} &\multicolumn{4}{c}{\phantom{(((}$\mathcal{F}=10$ J/cm$^2$} & \hspace{0.5cm} & \multicolumn{4}{c}{\phantom{((}$\mathcal{F}=20$ J/cm$^2$} & \hspace{0.5cm} & \multicolumn{4}{c}{\phantom{((}$\mathcal{F}=50$ J/cm$^2$}  \\ 
&&&& &\hspace{0.5cm} &\multicolumn{4}{c}{($\Omega=8$ MHz)} & & \multicolumn{4}{c}{($\Omega=11$~MHz)} & & \multicolumn{4}{c}{($\Omega=18$~MHz)}  \\ 
\cline{7-10}  \cline{12-15}  \cline{17-20} \\[-1.5ex]
$p$      & $T$    &  {$\Delta_l$}~ & $\Gamma_c$  & $\sigma_D$ && $\hat{\Gamma}_V$ &  $\Gamma_V$  & $\hat{\rho}_{33}$ & $\bar{\rho}_{33}$ & & $\hat{\Gamma}_V$ &  $\Gamma_V$   & $\hat{\rho}_{33}$ & $\bar{\rho}_{33}$ & & $\hat{\Gamma}_V$ &  $\Gamma_V$  & $\hat{\rho}_{33}$ & $\bar{\rho}_{33}$ \\
 $[\mathrm{bar}]$ & [K]    & [MHz]       & [MHz]       & [MHz]    &   & [MHz]      & [MHz] & & & & [MHz]      & [MHz]  &  & & & [MHz]      & [MHz]  \\ 
  [+1.0ex]  \hline	\\ [-1.0ex] 
  0.5            & 22     &  100      &    {770}       &    60  & &  {220}   &  {228}  &  {0.23}  &  {0.19}  & &  {223} &  {232} &  {0.45} &  {0.33} & &   {231} &  {267} & - & 
 {0.59} \\ 
  1              & 22     &  100      &    {913}       &    60   & &  {235}   &  {242}  &  {0.21}  &  {0.18}  &  &  {236} &  {256} &  {0.41} &  {0.32} & &  {241} &  {283} & - &  {0.60} \\ 
  0.5            & 22     &  10       &    {205}       &    60  & &  {161}   &  {167}  &  {0.36}  &  {0.25}  &&  {163} &  {176} &  {0.71} &  {0.38}    & &  {169} &  {190} & - &   {0.57}         \\      
  1              & 22     &  10        &    {348}       &    60  &  &  {174}   &  {183}   &  {0.31} &  {0.25}  &&  {175} &  {187} &  {0.63} &  {0.40} & &  {178} &  {208} & -&  {0.64} \\[1mm] 
  0.5            & 30     &  100       &    {733}       &    70 &  &  {239}   &  {247}   &  {0.21}  &  {0.17}  &&  {243} &  {251} &  {0.43} &  {0.30} &&  {254} &  {277} & - &   {0.54 }\\ 
  1              & 30     &  100       &    {838}       &    70 &  &  {249}   &  {256}   &  {0.20}  &  {0.17}  &&  {251} &  {269} &  {0.40} &  {0.31} & &  {257} &  {287} & - &  {0.57} \\ 
  0.5            & 30     &  10        &    {168}       &    70 &  &  {182}   &  {189}   &  {0.32}  &  {0.21}  &&  {185} &  {194} &  {0.65} &  {0.32} &&  {191} &  {208} & - &  {0.48} \\ 
  1              & 30     &  10        &    {273}       &    70 &  &  {190}   &  {199}   &  {0.30}  &  {0.22}  &&   {192} &  {204}  &  {0.60} &  {0.36} &&  {196} &  {226} & - &   {0.56} \\[1mm] 
  0.5            & 50     &  100       &    {693}       &    90 &  &  {283}   &  {288}   &  {0.19} &  {0.15}  &&  {290} &  {295} &  {0.38} &  {0.28} & &  {308} &  {316} &  {0.95} &   {0.42} \\ 
  1              & 50     &  100       &    {757}      &    90 &  &  {287}   &  {294}   &  {0.18}  &  {0.15}  &&   {290} &  {299} &  {0.37} &  {0.27} &&  {300} &  {328} &  {0.91} &  {0.49} \\ 
  0.5            & 50     &  10        &    {127}        &    90 &  &  {228}   &  {230}   &  {0.27}  &  {0.16}  &&  {233} &  {236} &  {0.53} &  {0.23} & &   {240} &  {249} & - &  {0.35} \\ 
 1              & 50     &  10        &    {192}       &    90 &  &  {231}   &  {236}  &  {0.26} &  {0.18}  &&  {231} &  {240} &  {0.51} &  {0.27} & &  {240} &  {256} & - &  {0.43} \\[1mm] 
  
 \\[-2.0ex]  \hline  \hline  \\[-2.0ex]
\end{tabular}
}
  \label{tab:widths} 
\end{table*}


%

\section{Conclusion}  
\label{sec:sum}

In view of the upcoming HFS experiment of the CREMA collaboration, we
calculated the combined probability that a thermalized  $\mu$p atom
undergoes laser excitation from the singlet to the triplet states followed by a
collisional deexcitation, where it acquires on average about 0.1~eV extra kinetic
energy.
This calculation was performed accounting for collisional and Doppler
effects.
The collisional effects together with the laser properties such as
bandwidth, exposure time, and fluence have been accounted for directly
in the three-level Bloch equations, while the Doppler broadening has been
accounted for in a second step by convoluting the results from the
Bloch equations with the Doppler profile.

We  derived simple analytical expressions valid in
limiting regimes (Fermi-golden rule and Rabi-oscillations).
Given the small Rabi frequency  relative to the homogeneous broadening, the proposed HFS experiment lies more closely to the Fermi-golden-rule regime.

The collisional effects play an important role: while the inelastic
collisions {trigger} a successful laser transition, 
the elastic collisions decrease the laser excitation probability to the triplet state. 
In the experimental conditions of the CREMA HFS experiment, (around 0.5~bar,  22~K, $\Delta_l=100$~MHz) the 
 decoherence effects {decrease}  the transition
probability for small flueces  by about a factor of  {two (see Table~\ref{tab:flue_E1}).}

For fluences reached in the optical cavity (up to 20 J/cm$^2$ locally), the  {
 dynamical effects beyond the Fermi approximation} captured by the numerical integration
of the Bloch equations are significant, as can be seen by comparing the solid and
dashed lines in Figure~\ref{fig:compare-analytical-numerical-Fermi-regime} and  {$\hat{\rho}_{33}$ and $\bar{\rho}_{33}$ in Table~\ref{tab:widths}. We also demonstrate that $\bar{\rho}_{33}$ is not significantly affected  by the assumed distribution of the hydrogen rotational states.}

%
{For the  a fluence of 10~J/cm$^2$, 100~MHz laser bandwidth and 100~ns laser exposure,    {19\%} of the $\mu$p atoms exposed to the laser light acquires the extra 0.1~eV kinetic energy for a hydrogen gas target at 22~K temperature and 0.5~bar pressure.   }

This paper is the first in a set of studies dedicated to the $\mu$p HFS experiment  {of the CREMA collaboration}.
In conjunction with studies of the $\mu$p {diffusion in the target gas, the} optical multi-pass
cavity and the detection system which will be published elsewhere, this
{allows to optimize} the hydrogen target and {define the specifications for} the laser system.




%
\section*{Acknowledgements}

We acknowledge the support of the following grants: Funda\c{c}\~{a}o
para a Ci\^{e}ncia e a Tecnologia (FCT), Portugal, and FEDER through
COMPETE in the framework of project numbers PTDC/FIS-AQM/29611/2017 and UID/04559/2020 (LIBPhys), and contracts No. \newline SFRH/BPD/92329/2013   and No. PD/BD/128324/2017; 
Deutsche Forschungsgemeinschaft (DFG, German Research Foundation) under
Germany's Excellence Initiative EXC 1098 PRISMA (194673446), 
Excellence Strategy EXC PRISMA+ (390831469) {and DFG/ANR Project LASIMUS (DFG Grant Agreement 407008443)}; 
{The French National Research Agency with project ANR-18-CE92-0030-02;} 
The European Research Council (ERC) through CoG. \#725039, and the Swiss National Science Foundation through the projects SNF 200021\_165854 and  SNF \newline 200020\_197052.

\appendix
\section{Hyperfine dipole matrix elements}
\label{sec:appen}

Evaluation of the matrix elements follows similar methods as done in recent works ~\cite{Safari2012, Amaro2015a, Amaro2015b, Amaro2018}. The hyperfine  matrix elements of equation~\eqref{eq:mar_E1M1} are evaluated with standard angular reduction methods, which start by expanding the electric dipole  $\bm{r}\cdot \bm{\hat{\varepsilon}} $, and the magnetic dipole $ \bm{S} \cdot\bm{\hat{k}\times \hat{\varepsilon} }$ operators (similarly with  $ \bm{I} \cdot\bm{\hat{k}\times \hat{\varepsilon} }$)  in a spherical basis. After  considering the overall atomic state being the product coupling of the nucleus and electron angular momenta, and using the Wigner-Eckart theorem,  the matrix element for  the  irreducible rank-one tensor $T_\lambda$ (components $\lambda=\pm1$) of the spherical expansion is given by \cite{Johnson2007,Varshalovich1988}

%
\begin{eqnarray}\left\langle \beta' F' m' J'\left|T_{\lambda}\right| \beta F m J\right\rangle &=& 
 (-1)^{F'+I+F+1+J'-m'}  \sqrt{[F, F']}\left(\begin{array}{ccc}
F' & 1 & F \nonumber  \\
-m' & \lambda & m\end{array}\right) \\ &\times&
\left\{ \begin{array}{ccc}
J & I & F\\
F' & 1 & J'\end{array}\right\} \left\langle \beta' J'\|T\|\beta J\right\rangle ~.
\label{eq:tota_redu_jj}
\end{eqnarray}
%
%
Here,  $T_{\lambda}=r_\lambda$ and $T_{\lambda}=S_\lambda$ are for the $2s-2p$ and HFS cases, respectively. Both final and initial states are defined by $\left| f \right\rangle=\left| \beta' F' m' J' \right\rangle$ and $\left| i \right\rangle=\left|  \beta F m J \right\rangle$, where $F$ is the total angular momentum , $m$ is its projection, and $J$ is the total  angular momentum of $\mu$. $\beta$ represents further quantum numbers, such as  $\ell$ and $n$. The notation $[j_1, j_2,...]$ stands for  $(2j_1+1)(2j_2+1)$... The  reduced matrix elements for E1 and M1 are given by

\begin{eqnarray}
 \left\langle 2pJ' \| r \| 2s,J=1/2\right\rangle  &=& \sqrt{\left[1/2, J' \right]}
  \left(\begin{array}{ccc}
J' & 1 & 1/2\\
1/2 & 0 & -1/2\end{array}\right) \int P_{2p}' \,P_{2s}\,rdr ~, 
\label{eq:redu_jj_E1}
\end{eqnarray}
and 
\begin{eqnarray}
 \left\langle  \beta' J' | S  | \beta J\right\rangle  =\frac{ \sqrt{6}}{2}\delta( \ell, \ell')(-1)^{J'-1/2}\sqrt{[J,J']}  
 \left\{ \begin{array}{ccc}
1/2 & \ell & J \\
J' & 1 & 1/2
\end{array}\right\} \hbar~,
\label{eq:tota_redu_jj_M1}
\end{eqnarray}
for the $\bm{r}$ and $\bm{S}$ operators, respectively. The functions $P_{2p}'$ and $P_{2s}$ in equation~\eqref{eq:redu_jj_E1} are hydrogenic radial wavefunctions.  

After performing the averaging over initial magnetic substates $m$ and  summation over final substates $m'$, the  matrix element are given by 
%
%
%
\begin{eqnarray}
{\mathcal{M}}_{\mbox{E1}} &=&  \frac{1}{\sqrt{3}} \sqrt{[F', J, J']} \left\{ \begin{array}{ccc}
J & I & F\\
F' & 1 & J'\end{array}\right\}   \left(\begin{array}{ccc}
J' & 1 & J\\
1/2 & 0 & -1/2\end{array}\right)   \int P'_{2p}\,P_{2s}\,rdr  ~, \label{eq:E1_expli}\\
 {\mathcal{M}}_{\mbox{M1}}  &=& \frac{1}{\sqrt{8}}  \sqrt{[F',F]}    \left\{ \begin{array}{ccc}
1/2 & I & F\\
F'& 1 & 1/2\end{array}\right\}  
 \left(\frac{g_\mu \hbar}{m_\mu c}  + \frac{g_p \hbar}{m_p c} \right) ~,
 \label{eq:M1_expli}
\end{eqnarray}
for the  Lamb shift and HFS, respectively. While evaluating the matrix elements,
the overall contribution of light polarization is reduced to
$|\hat{\varepsilon} |^2=1$, thus suppressing any dependence on the
laser polarization (circular or linear) used. This is expected since the
initial state is not polarized, and  the final state polarization
is not observed. Equation~\eqref{eq:M1_expli} is equivalent to  equation~(2) of Ref.~\cite{MohanDas2020} with $g_\mu=2$,  $g_p=0$ and $1/c\rightarrow e$ (different units), however we obtain a value of $e\hbar /{(2 mc)}$ instead of  $e\hbar/ (2 \sqrt{6} mc)$ after its evaluation.

\section{Results for a statistical distribution and $\tau=10$~ns}
\label{sec:Appenstatistic}

Table \ref{tab:flue_E1_stat} compares $\mathcal{F}_{\text{sat}}$ and $\Gamma_c$  for statistical and Boltzmann distributions between ortho- and para-hydrogen, while Table~\ref{tab:widths_sta} lists the transition probabilities $\bar{\rho}_{33}(\omega_r)$ and linewidths, similar to Table~\ref{tab:widths} but for a statistical  distribution.  Differences of less than   {5\%} are observed for both the saturation fluences and $\bar{\rho}_{33}(\omega_r)$ at the listed experimental conditions.

 Table~\ref{tab:widths_time} contains the transition probabilities $\bar{\rho}_{33}(\omega_r)$ and linewidths for $\tau=10$~ns.  As observed,  differences in $\bar{\rho}_{33}(\omega_r)$ between 10~ns and 100~ns are more pronounced for $\Delta_l=100$~MHz, having differences around 26\%,  while for $\Delta_l=10$~MHz are negligible. 

 \begin{table}
 \centering
 \caption{ Decoherence rates and saturation fluences for statistical and Boltzmann distributions between ortho- and para-hydrogen, and for $\Delta_l=100$~MHz.
 %
    }
 {
\begin{tabular} {cccccc} 
\\[-2.0ex]  \hline  \hline  \\[-2.0ex]
$T$   & $p$    & $\Gamma_c^{\text{Boltz.} }$  & $\Gamma_c^\text{Stat.}$ & $\mathcal{F}^{\text{Boltz.} }_{\mbox{sat}}$ & $\mathcal{F}_{\mbox{sat}}^{\text{Stat.} } $  \\ 
~[K] & [bar]   & [MHz]  & [MHz] & [J/cm$^2$] & [J/cm$^2$] \\
  [+1.0ex]  \hline	\\ [-1.0ex]
        22   & 0.5             &      771     & 782    & 44 & 45\\ 
        22   & 1                 &     913    & 936   & 49 & 49\\ 
        22   & 2                  &   1198      & 1244    & 58 & 59\\[1mm]
        30   & 0.5                &   733    & 743   & 47 & 47\\
        30   & 1                    &  838      & 859     & 50 & 50\\[1mm]
        50   & 0.5                  &  693    & 699     & 53 & 53\\
        50   & 1              &     757      & 770     & 55 & 55\\ 
   
 \\[-2.0ex]  \hline  \hline  \\[-2.0ex]
\end{tabular}}
  \label{tab:flue_E1_stat} 
\end{table}

\begin{table*}
   \caption{ Same as Table~\ref{tab:widths} but for a statistical ortho- to para-distribution of the initial rotational states of H$_2$ molecules.}
   \resizebox{\textwidth}{!}{
\begin{tabular} {cccccccccccccccccccc} 
\\[-2.0ex]  \hline  \hline  \\[-2.0ex]
&&&& &\hspace{0.5cm} &\multicolumn{4}{c}{\phantom{(((}$\mathcal{F}=10$ J/cm$^2$} & \hspace{0.5cm} & \multicolumn{4}{c}{\phantom{((}$\mathcal{F}=20$ J/cm$^2$} & \hspace{0.5cm} & \multicolumn{4}{c}{\phantom{((}$\mathcal{F}=50$ J/cm$^2$}  \\ 
&&&& &\hspace{0.5cm} &\multicolumn{4}{c}{($\Omega=8$ MHz)} & & \multicolumn{4}{c}{($\Omega=11$~MHz)} & & \multicolumn{4}{c}{($\Omega=18$~MHz)}  \\ 
\cline{7-10}  \cline{12-15}  \cline{17-20} \\[-1.5ex]
$p$      & $T$    &  {$\Delta_l$}~ & $\Gamma_c$  & $\sigma_D$ && $\hat{\Gamma}_V$ &  $\Gamma_V$  & $\hat{\rho}_{33}$ & $\bar{\rho}_{33}$ & & $\hat{\Gamma}_V$ &  $\Gamma_V$   & $\hat{\rho}_{33}$ & $\bar{\rho}_{33}$ & & $\hat{\Gamma}_V$ &  $\Gamma_V$  & $\hat{\rho}_{33}$ & $\bar{\rho}_{33}$ \\
 $[\mathrm{bar}]$ & [K]    & [MHz]       & [MHz]       & [MHz]    &   & [MHz]      & [MHz] & & & & [MHz]      & [MHz]  &  & & & [MHz]      & [MHz]  \\ 
  [+1.0ex]  \hline	\\ [-1.0ex] 
  0.5            & 22     &  100       &    {782}       &    60  & &  {222} &  {231} &  {0.22} &  {0.18} & &   {225} &  {234} &  {0.45} &  {0.32} & &  {235} &  {269} & - &  {0.58}  \\ 
  1              & 22     &  100       &    {936}       &    60   & &  {238} &  {246} &  {0.20} &  {0.18}   &  &  {240} &  {259} &  {0.40} &  {0.32} & &  {245} &  {287} & - &  {0.59}  \\ 
  0.5            & 22     &  10        &    {217}       &    60  & &  {162} &  {169} &  {0.35} &  {0.25}  &&  {165} &  {177} &  {0.71} &  {0.38}    & & {171} &  {193} & - &  {0.57}        \\      
  1              & 22     &  10        &    {371}      &    60  &  &  {176} &  {182} &  {0.31} &  {0.24}   &&  {177} &  {191} &  {0.62} &  {0.40} & &  {182} &  {213} & - &  {0.64} \\[1mm] 
  0.5            & 30     &  100       &    {744}      &    70 &  &  {241} &  {249} &  {0.21} &  {0.17}  &&  {245} &  {253} &  {0.43} &  {0.29}   &&  {258} &  {279} & - &  {0.53} \\ 
  1              & 30     &  100       &    {859}       &    70 &  &  {251} &  {259} &  {0.20} &  {0.17}  &&   {254} &  {270} &  {0.40} &  {0.30}  & &  {261} &  {290} &  {0.99} &  {0.56}  \\ 
  0.5            & 30     &  10        &    {178}       &    70 &  &  {183} &  {190} &  {0.32} &  {0.21}  &&  {186} &  {196} &  {0.64} &  {0.32}  &&  {194} &  {211} & - &  {0.48} \\ 
  1              & 30     &  10        &    {293}       &    70 &  &  {192} &  {197} &  {0.29} &  {0.22}  &&   {194} &  {208} &  {0.59} &  {0.36}  &&  {199} &  {226} & - &  {0.56} \\[1mm] 
 0.5            & 50     &  100       &    {699}      &    90 &  &  {285} &  {290} &  {0.19} &  {0.13}  &&   {292} &  {295} &  {0.38} &  {0.23} & &  {310} &  {317} &  {0.95} &  {0.41} \\ 
 1              & 50     &  100       &    {770}       &    90 &  &  {288} &  {295} &  {0.18} &  {0.15}   &&   {292} &  {301} &  {0.36} &   {0.26}  &&  {303} &  {331} &  {0.91} &  {0.48} \\ 
 0.5            & 50     &  10        &    {134}       &    90 &  &   {229} &  {233} &  {0.27} &  {0.16}  &&   {233} &  {234} &  {0.53} &  {0.23}  & &  {242} &  {251} & - &  {0.35} \\ 
 1              & 50     &  10        &    {205}       &    90 &  &  {233} &  {239} &  {0.25} &  {0.18} &&  {235} &  {243} &  {0.51} &  {0.28} & &   {242} &  {260} & - &  {0.43} \\[1mm] 

 \\[-2.0ex]  \hline  \hline  \\[-2.0ex]
\end{tabular}
}
  \label{tab:widths_sta} 
\end{table*}

\begin{table*}
   \caption{Same as Table~\ref{tab:widths} but for a laser  exposure time of $\tau=10$~ns. }
   \resizebox{\textwidth}{!}{
    {
\begin{tabular} {cccccccccccccccccccc} 
\\[-2.0ex]  \hline  \hline  \\[-2.0ex]
&&&& &\hspace{0.5cm} &\multicolumn{4}{c}{\phantom{(((}$\mathcal{F}=10$ J/cm$^2$} & \hspace{0.5cm} & \multicolumn{4}{c}{\phantom{((}$\mathcal{F}=20$ J/cm$^2$} & \hspace{0.5cm} & \multicolumn{4}{c}{\phantom{((}$\mathcal{F}=50$ J/cm$^2$}  \\ 
&&&& &\hspace{0.5cm} &\multicolumn{4}{c}{($\Omega=26$ MHz)} & & \multicolumn{4}{c}{($\Omega=36$~MHz)} & & \multicolumn{4}{c}{($\Omega=58$~MHz)}  \\ 
\cline{7-10}  \cline{12-15}  \cline{17-20} \\[-1.5ex]
$p$      & $T$    &  {$\Delta_l$}~ & $\Gamma_c$  & $\sigma_D$ && $\hat{\Gamma}_V$ &  $\Gamma_V$  & $\hat{\rho}_{33}$ & $\bar{\rho}_{33}$ & & $\hat{\Gamma}_V$ &  $\Gamma_V$   & $\hat{\rho}_{33}$ & $\bar{\rho}_{33}$ & & $\hat{\Gamma}_V$ &  $\Gamma_V$  & $\hat{\rho}_{33}$ & $\bar{\rho}_{33}$ \\
 $[\mathrm{bar}]$ & [K]    & [MHz]       & [MHz]       & [MHz]    &   & [MHz]      & [MHz] & & & & [MHz]      & [MHz]  &  & & & [MHz]      & [MHz]  \\ 
  [+1.0ex]  \hline	\\ [-1.0ex] 
  0.5            & 22     &  100      &   {770}       &    60  & & {245}   & {257}  & {0.23}  & {0.14}  & & {269} & {258} & {0.45} & {0.24} & &  {328} & {291} & - & {0.42} \\ 
  1              & 22     &  100      &   {913}       &    60   & & {249}   & {271}  & {0.21}  & {0.15}  &  & {263} & {282} & {0.41} & {0.27} & & {301} & {304} & - & {0.49} \\ 
  0.5            & 22     &  10       &   {205}       &    60  & & {177}   & {189}  & {0.36}  & {0.24}  && {190} & {190} & {0.72} & {0.39}    & & {219} & {208} & - &  {0.56}         \\      
  1              & 22     &  10        &   {348}       &    60  &  & {184}   & {205}   & {0.31} & {0.22}  && {194} & {210} & {0.63} & {0.38} & & {218} & {227} & -& {0.60} \\[1mm] 
  0.5            & 30     &  100       &   {733}       &    70 &  & {271}   & {276}   & {0.21}  & {0.11}  && {301} & {278} & {0.43} & {0.20} && {371} & {312} & - &  { 0.34}\\ 
  1              & 30     &  100       &   {838}       &    70 &  & {267}   & {285}   & {0.20}  & {0.14}  && {285} & {285} & {0.40} & {0.25} & & {330} & {319} & - & {0.44} \\ 
  0.5            & 30     &  10        &   {168}       &    70 &  & {201}   & {207}   & {0.33}  & {0.22}  && {214} & {208} & {0.65} & {0.36} && {246} & {226} & - & {0.49} \\ 
  1              & 30     &  10        &   {273}       &    70 &  & {203}   & {215}   & {0.30}  & {0.21}  &&  {214} & {221}  & {0.60} & {0.35} && {240} & {238} & - &  {0.54} \\[1mm] 
  0.5            & 50     &  100       &   {693}       &    90 &  & {333}   & {314}   & {0.19} & {0.08}  && {374} & { 315} & {0.38} & {0.13} & & {470} & {351} & {0.95} &  {0.22} \\ 
  1              & 50     &  100       &   {757}      &    90 &  & {315}   & {315}   & {0.18}  & {0.10}  &&  {341} & {327} & {0.37} & {0.19} && {404} & {350} & {0.91} & {0.33} \\ 
  0.5            & 50     &  10        &   {127}        &    90 &  & {252}   & {246}   & {0.27}  & {0.18}  && {269} & {253} & {0.53} & {0.29} & &  {305} & {272} & - & {0.38} \\ 
 1              & 50     &  10        &   {192}       &    90 &  & {249}   & {252}  & {0.26} & {0.18}  && {263} & { 258} & {0.51} & {0.30} & & {293} & {285} & - & {0.43} \\[1mm] 
  
 \\[-2.0ex]  \hline  \hline  \\[-2.0ex]
\end{tabular}
}
}
  \label{tab:widths_time} 
\end{table*}

\bibliography{HFS_articles_v1}




\nolinenumbers

\end{document}